\documentclass[aps,pra,floatfix,tightenlines,twocolumn,amsmath,amssymb]{revtex4}
\pdfoutput=1

\usepackage{amsmath,bbm,graphicx}
\usepackage[justification=justified]{subfig}
\usepackage[letterpaper=true,pdftex,pdfpagelabels=true,bookmarks = true,colorlinks = true, citecolor=blue, urlcolor=cyan]{hyperref}

\newcommand{\bra}[1]{\ensuremath{\left\langle #1\right\vert}}
\newcommand{\ket}[1]{\ensuremath{\left\vert #1\right\rangle}}
\newcommand{\expval}[1]{\ensuremath{\left\langle #1 \right\rangle}}
\newcommand{\hsp}[1]{\hspace{#1 em}}
\newcommand{\sqz}{\hsp{-0.1}}
\newcommand{\ketbra}[2]{\left\vert{#1}\right\rangle \sqz\sqz\sqz \left\langle{#2}\right\vert}
\newcommand{\braket}[2]{\left\langle{#1}\right\vert \sqz \sqz \left. {#2}\right\rangle}

\newcommand{\tr}{\text{Tr}}
\def\iden{\mathbbmss{1}}

\graphicspath{{Figures/}}

\begin{document}
\title{Quantum Throughput: Quantifying quantum communication devices with homodyne measurements}
\date{\today}

\author{N. Killoran$^1$, H. H\"{a}seler$^{1,2}$, and N. L\"{u}tkenhaus$^{1,2}$}
\affiliation{$^1$Institute for Quantum Computing and Department of Physics \&
Astronomy, University of Waterloo, N2L 3G1 Waterloo, Canada\\
$^2$Max Planck Institute for the Physics of Light, Universit\"{a}t Erlangen-N\"{u}rnberg, 91058 Erlangen, Germany}

\begin{abstract}
Quantum communication relies on optical implementations of channels, memories and repeaters. In the absence of perfect devices, a minimum requirement on real-world devices is that they preserve quantum correlations, meaning that they have some thoughput of a quantum mechanical nature. Previous work has verified throughput in optical devices while using minimal resources. We extend this approach to the quantitative regime. Our method is illustrated in a setting where the input consists of two coherent states while the output is measured by two homodyne measurement settings.
\end{abstract}

\maketitle

\section{Introduction}\label{sec:intro}

The introduction of new quantum mechanical technologies promises to fundamentally alter the way we communicate. Quantum key distribution (QKD), for instance, will allow us to communicate in an intrinsically secure way \cite{bennett84a, scarani08a}. But new quantum communication technologies will require a new telecommunications infrastructure, one which is quantum-enabled. That is, this network must be able to properly accommodate the quantum properties that quantum communications inherently rely on. Such a quantum network will contain many novel components, such as quantum memories \cite{lvovsky09a}, quantum repeaters \cite{briegel98a}, or, most generally, quantum channels. These components must each operate in a strictly quantum way. 

Of course, no technology is perfect, and quantum technologies offer a new set of practical challenges. However, as we have learned from QKD, perfectly ideal devices are not a necessity. By shifting our efforts into classical post-processing of data, we can deal with imperfections in quantum technologies. The question then becomes, how much imperfection can be tolerated before a device is no longer operating in a sufficiently quantum way? We can enforce a minimal quantum requirement on devices by insisting that they do not act as \emph{measure and prepare} channels \cite{haseler08a} (or, in the parlance of QKD, \emph{intercept and resend} channels), since communication through such channels is equivalent to classical communication. Indeed, this type of channel destroys any quantum correlations in bipartite states when one subsystem is sent through it. 

Of course, this is just the minimum requirement. It is also important to quantify the quantum behaviour, as is done in the field of entanglement measures, or in QKD through the secret key rate. For quantum channels, we can ask, \emph{how well does the channel preserve quantum correlations in bipartite systems, when only one subsystem passes through it?} To study this question, we take a state with well-quantified quantum correlations, send one subsystem through the channel, and examine the output. We then compare the quantum correlations detectable in the output with the input correlations. In fact, as we shall see, we can test for these correlations in a so-called `prepare and measure' picture, bypassing the need to use actual bipartite states. A strong quantum channel is one which preserves all or nearly all of the quantum correlations. This idea corresponds to what we shall call the \emph{quantum throughput}. Such a measure would allow us to characterize the suitability of devices for quantum communication tasks. The goal of this work is to illustrate that these ideas about device characterization via quantum throughput can be implemented in a meaningful way. Although we will make specific choices regarding device types or quantification measures, the basic idea remains quite general, and our scheme can be extended and adapted to other methods as well.

Finally, if we picture a future quantum communications network consisting of many components, it should be evident that any device-testing procedure should be as experimentally practical as possible. Ideally, we seek a testing scenario where a finite number of test states and a limited set of measurements are sufficient to understand the quantum throughput. The latter requirement is especially important for optical systems, which are perhaps the most natural choice of carrier for quantum information. In these systems, full tomography is not really a practical option because of the dimension of the Hilbert space. We have previously examined quantum correlations in optical devices in a qualitative way \cite{haseler08a}; in the present contribution, we will extend those results to provide a quantitative picture of optical devices.

The rest of this paper is organized as follows. In Sec. \ref{sec:quant} we outline our quantitative device-testing scheme, focusing mainly on optical systems. We show how to estimate important parameters from homodyne measurements on the output, and how to use these estimates to make quantitative statements about the optical device. In Sec. \ref{sec:results}, we give the results of this quantification procedure for a wide class of optical channels, and examine the strength of our method. Sec. \ref{sec:conclusion} summarizes the paper, while Appendices \ref{app:overlapbounds}-\ref{app:offdiagbounds} provide technical details and derivations.

\section{Quantification procedure}\label{sec:quant}
\subsection{Device testing scenario}\label{ssec:quantscheme}

The quantum device testing procedure we employ is the same as the one found in \cite{haseler08a}. This protocol is based on the idea that a truly quantum channel should be distinguishable from those channels where the input quantum state is temporarily converted to classical data before a new quantum state is output, a so-called \emph{measure and prepare} channel. Measure and prepare channels are also called \emph{entanglement-breaking} channels, as the two notions are equivalent \cite{horodecki03a}. This provides a hint on how to quantify a channel's quantum throughput, namely by sending part of an entangled state through the channel and determining the amount of entanglement that still remains afterwards. 

To this end, imagine we have an entangled state of the form
\begin{equation}
\label{eq:initialstate}
     \ket{\psi}_{AB} = \frac{1}{\sqrt{2}}\left[ \ket{0}_A\ket{\alpha}_B + \ket{1}_A\ket{-\alpha}_B \right]
\end{equation}
where system $A$ is a qubit and system $B$ is an optical mode. We can assume, without loss of generality, that $\alpha\in \mathbb{R}$, so that $\ket{\alpha}$ and $\ket{-\alpha}$ denote coherent states of opposite phase. This is an entangled state for all values $\alpha\neq 0$, as can be seen by calculating the entropy of entanglement. Keeping subsystem A isolated, an optical channel can be probed using subsystem B of this state, followed by local projective measurements $\{\ket{0}\bra{0}, \ket{1}\bra{1}\}$ by Alice and homodyne measurements $\{\hat{x},\hat{p}\}$ by Bob. These expectation values, along with the knowledge of Alice's reduced density matrix $\rho_A$, can be used to determine just how much of the initial state's entanglement is remaining. 

Of course, states like Eq. (\ref{eq:initialstate}) may be difficult to create and therefore not suited for practical device testing. However, notice that Alice's reduced density matrix does not depend on what happens in the optical channel, nor on any of Bob's measurement results. Her expectation values can be completely determined from the initial state $\ket{\psi}_{AB}$. Indeed, Alice's measurement results can be thought of as classical registers which merely record which mode state was sent through the device. This observation allows us to move from an entanglement-based (EB) picture to an equivalent `prepare and measure' (PM) scenario \cite{bennett92c,bennett92b}, in which Alice's measurements are absorbed into the initial state preparation. 

In a PM scenario, we retain full knowledge of $\rho_A$, in particular the off-diagonal coherence term $\left[\rho_A\right]_{01}=\braket{\alpha}{-\alpha}$. We must insert this additional information \emph{by hand} into the set of expectation values for $\rho_{AB}$. This distinguishes the expectation values from data which would come from using just a classical mixture of test states $\rho_{AB}^{\text{cl.}}=\frac{1}{2}\left(\ket{\alpha}\bra{\alpha}+\ket{-\alpha}\bra{-\alpha}\right)$. Other than this, the procedure is the same as the EB scenario described above. Quantum correlations introduced in this way are referred to as `effective entanglement' \footnote{By `effective entanglement', we mean quantum correlations in the PM picture which translate to entanglement in the EB picture. This should not be confused with the usage in \cite{roszak10a}, where the term refers to the minimal entanglement compatible with the given measurement results. Since we will make use of both ideas in this work, we will refer to the latter simply as the `minimal entanglement.'}. Using this convenient theoretical trick, the testing protocol can be accomplished simply by probing the channel using a source which prepares one of the two conditional states $\{\ket{\alpha}, \ket{-\alpha} \}$ with equal probability. If the measured expectation values, along with the inserted knowledge of $\rho_A$, are not compatible with any separable qubit-mode state, then there is (effective) entanglement and the channel is certifiably quantum. Exploiting the duality between the PM picture and the EB picture, we can quantify the quantum correlations remaining in the output state through a suitable entanglement measure. In turn, this can be compared to the entanglement of the state in Eq. (\ref{eq:initialstate}) to determine the quantum throughput.

\subsection{Quantification scheme}
Our main goal in this work is to give an estimate of the amount of effective entanglement observable in an optical system after transmission through an optical channel. Our method is based on the following observation: When the two, initially pure, conditional states $\{\ket{\alpha},\ket{-\alpha}\}$ pass through the channel, they are subject to loss and noise, and evolve in general to mixed states $\{\rho_0,\rho_1 \}$ on the infinite-dimensional mode Hilbert space; however, since we work with coherent states, this change in purity comes only from the noise. Thus, for any loss value, if the noise introduced by the channel is not too high, then the output states $\rho_0$ and $\rho_1$ will still be nearly pure. In this case, most of the information about the state is still contained in a very small subspace of the full infinite dimensional Hilbert space. Estimating the `most significant' subspace on the mode system can therefore be quite useful. This subspace should contain as much information as possible about both conditional states. Additionally, we will concentrate on the simplest non-trivial mode subspace, namely one of dimension 2. Writing the conditional output states $\rho_0$ and $\rho_1$ in terms of their eigenvectors, in order of descending eigenvalues, we have
\begin{equation}
     \rho_j = \sum_{k=0}^{\infty} \lambda_j^k\ketbra{\lambda_j^k}{\lambda_j^k}.
     \label{eq:eigendecomp}
\end{equation}
The most significant subspace is then the one formed using $\ket{\lambda_0^0}$ and $\ket{\lambda_1^0}$ as basis vectors. 

Three parameters will be important to identify this subspace: $\lambda_0^0$, $\lambda_1^0$, and $\braket{\lambda_0^0}{\lambda_1^0}$. We will estimate these parameters using homodyne detection. Specifically, if $\hat{a}$ is the annihilation operator for the mode at Bob's detector, then a balanced homodyne detection scheme allows us to measure the field quadratures, here defined as
\begin{equation}
     \hat{x} = \frac{1}{\sqrt{2}}\left(\hat{a}^\dagger +  \hat{a}\right), ~ \hat{p} = \frac{i}{\sqrt{2}}\left(\hat{a}^\dagger - \hat{a} \right).
\end{equation}
We will use the mean values $\expval{\hat{x}}_{0/1},\expval{\hat{p}}_{0/1}$ and the variances 
\begin{align}
     V_{0/1}(\hat{x}) = \expval{\hat{x}^2}_{0/1}-\expval{\hat{x}}_{0/1}^2&,\label{eq:variance1} \\
     V_{0/1}(\hat{p}) = \expval{\hat{p}^2}_{0/1}-\expval{\hat{p}}_{0/1}^2&\label{eq:variance2}
\end{align}
of the quadratures from both conditional states to estimate the three subspace parameters. Exactly how this is done will be shown in the next part. With these parameters, we can build a $4\times 4$ density matrix $\rho_P$, which corresponds to the projection of the full qubit-mode state $\rho_{AB}$ onto the two-qubit subspace spanned by the basis $\{\ket{0}_A,\ket{1}_A\}\otimes\{\ket{\lambda_0^0}_B,\ket{\lambda_1^0}_B\}$. 

The idea is now to bound the entanglement of the full state $\rho_{AB}$ using the entanglement of the projection $\rho_P$. For this, we need to exploit the strong monotonicity under local operations and classical communication (LOCC) property found in many entanglement measures. Specifically, if we perform a complete set of local measurements on a bipartite state $\sigma$, which yields (perhaps using some classical communication) the state $\sigma_m$ with probability $p_m$, then the strong monotonicity property captures the idea that the entanglement should not increase, on average, under this process. In other words, for a given measure $\mathcal{E}$,
\begin{equation}
     \mathcal{E}(\sigma) \geq \sum_m p_m\mathcal{E}(\sigma_m).
\end{equation}
As the name implies, this is a stronger condition than just monotonicity under LOCC alone. In our case, the measurement consists of projecting the mode system onto the most significant subspace or onto the orthogonal complement. We denote the former projection by 
\begin{equation}
     P = \iden_A\otimes\left( \ketbra{\lambda_0^0}{\lambda_0^0}_B + \ketbra{\lambda_1^0}{\lambda_1^0}_B\right)
\end{equation}
and the latter by $P^\perp = \iden_{AB}-P$. Then, if we choose an entanglement measure $\mathcal{E}$ with the strong monotonicity property, we have
\begin{equation}
     \mathcal{E}(\rho_{AB}) \geq p \mathcal{E}\left(\frac{P\rho_{AB} P}{p}\right) + q \mathcal{E}\left(\frac{P^\perp\rho_{AB} P^\perp}{q}\right),
     \label{eq:strongmono}
\end{equation}
where $p=\tr(P\rho_{AB} P)$ and $q=\tr(P^\perp\rho_{AB} P^\perp)$.

For later practicality purposes, we would like to factor the probabilities through the entanglement measure, so that we work directly with unnormalized states. The unnormalized projected state is thus given by $\rho_P = P\rho_{AB} P$. We must be careful to choose an entanglement measure which, in addition to being a strong monotone, can be defined for unnormalized states and which permits a positive prefactor to be absorbed into the state. We will focus on the Negativity \cite{zyczkowski98a,vidal02a,lee00a} in this work, for which this choice is justified.

We will not attempt to estimate the second term in Eq. (\ref{eq:strongmono}) coming from the orthogonal projection; we only note that it is non-negative, so that we have the bound
\begin{equation}
     \mathcal{E}(\rho_{AB}) \geq \mathcal{E}(\rho_P).
\end{equation}

In practice, the projected matrix $\rho_P$ will not be fully characterized and will contain open parameters. On the other hand, some constraints can be imposed on $\rho_P$ from our knowledge of the initial conditional states and the homodyne measurement results, as well as natural positivity ($\rho_P\geq 0$) and trace constraints $(\tr(\rho_P)\leq 1$ for unnormalized $\rho_P$). As a final step, we must determine the minimal entanglement of $\rho_P$ compatible with all allowed values of these open parameters, subject to the known constraints. We are left with the final relation
\begin{equation}
     \mathcal{E}(\rho_{AB}) \geq \min_{\rho_P}\mathcal{E}(\rho_P),
\end{equation}
which will be used as the basis for calculating bounds on $\mathcal{E}(\rho_{AB})$. The next two subsections will cover how to estimate $\rho_P$ and how we minimize the entanglement over all compatible forms of $\rho_P$.

\subsection{Estimating the projected state}\label{ssec:estimate}

The first step in our method requires determining the projection $\rho_P$ of the full state $\rho_{AB}$ onto the most-significant subspace. For this, we need to estimate the three parameters $\lambda_0^0$, $\lambda_1^0$ and $\braket{\lambda_0^0}{\lambda_1^0}$ from the decomposition in Eq. (\ref{eq:eigendecomp}). In Ref. \cite{rigas06b}, which considers the related problem of effective entanglement verification using heterodyne measurements (i.e. full knowledge of the $Q$ function), several useful formulas for estimating these maximal eigenvalues and overlaps are given. These bounds are later refined in \cite{zhao09a}, where they are used to derive secret key rates for continuous variable quantum key distribution. Here, we use these bounds as a starting point to build up a good estimate of the projected state for our quantification scheme. We will roughly follow the notation of \cite{zhao09a} in the following.

First, since the conditional output states $\{\rho_0,\rho_1\}$ have unit trace, their maximal eigenvalues can be parameterized by $\lambda_j^0 = 1-\tilde{\varepsilon}_j~(j=0,1)$, with $\tilde{\varepsilon}_j\in \left[0,1\right]$. Then Eqs. (66) and (69) from \cite{zhao09a} give directly the following bound: 
\begin{equation}\label{eq:noisebound}
 \tilde{\varepsilon}_j \leq \frac{1}{2}\left[\left(V_j(x) + \frac{1}{2} \right)\left(V_j(p) + \frac{1}{2} \right) -1 \right] =: U_j.
\end{equation}
This bound comes up several times, so it is denoted $U_j$ ($j=0,1$) to make later equations more readable. Importantly, the bound can be calculated using only the measured variances of the conditional states. 

Estimating the overlap $\braket{\lambda_0^0}{\lambda_1^0}$ is more involved. We need to derive bounds on its magnitude based on our available information. Again, we begin with bounds provided in Refs. \cite{rigas06b,zhao09a}. With suitable relaxations, their bounds can be put into a specific form which will be more desirable for us later, as we would ultimately like to do a convex optimization. The specific details of this relaxation are straightforward, and are outlined in Appendix \ref{app:overlapbounds}. We will need an additional parameter, $\kappa$, which can be calculated directly using the measured first moments $\{\expval{\hat{x}}_j,\expval{\hat{p}}_j\}$. Defining two coherent states with the same means as the conditional states,
\begin{equation}
 \ket{\beta_j} = \ket{\frac{1}{\sqrt{2}}\left(\expval{\hat{x}}_j + i\expval{\hat{p}}_j\right)},
 \label{eq:cohmean}
\end{equation}
the new parameter is given through the overlap of these coherent states, 
\begin{equation}
 \kappa := |\braket{\beta_0}{\beta_1}|.
\end{equation}
With this definition in place, we can give the relaxed bounds
\begin{equation}
 b_l(U_0,U_1, \kappa) \leq |\braket{\lambda_0^0}{\lambda_1^0}| \leq b_u(U_0,U_1,\kappa)
 \label{eq:overlapbounds}
\end{equation}
where
\begin{align}
      	b_l =  ~&\kappa\sqrt{1-2U_0}\sqrt{1-2U_1}
	       -\sqrt{1-\kappa^2}\sqrt{\frac{U_0}{1-2U_0}}\notag\\
	       -&\sqrt{1-\kappa^2}\sqrt{\frac{U_1}{1-2U_1}}	
	       -\sqrt{\frac{U_0}{1-2U_0}}\sqrt{\frac{U_1}{1-2U_1}}
	       \label{eq:blower}
\end{align}
and
\begin{align}
      	b_u = ~&\kappa + \sqrt{1-\kappa^2} \sqrt{\frac{U_0}{1-2U_0}}
	       + \sqrt{1-\kappa^2} \sqrt{\frac{U_1}{1-2U_1}}\notag\\
	       +&\sqrt{\frac{U_0}{1-2U_0}} \sqrt{\frac{U_1}{1-2U_1}}. 
	       \label{eq:bupper}
\end{align}

Having these bounds, obtained purely through homodyne measurements, we can now move on to estimating the elements of the projected density matrix $\rho_P$. We can already estimate matrix elements of the form $\bra{\lambda_j^0}\rho_j\ket{\lambda_j^0}$ using Eq. (\ref{eq:noisebound}), but to build $\rho_P$ we also require bounds on the supplementary elements $\bra{\lambda_i^0}\rho_j\ket{\lambda_i^0}$ for $i\neq j$. To get these, we first expand $\rho_0$ into its eigenbasis, Eq. (\ref{eq:eigendecomp}). Then, using the fact that $|\braket{\lambda_1^0}{\phi}|^2\in\left[0,1\right]$ for any normalized vector $\ket{\phi}$, we can easily derive the following bounds on the desired matrix element (see Appendix \ref{app:suppdiagbounds} for full details):
\begin{align}
     \label{eq:suppdiagbounds1}\bra{\lambda^0_1}\rho_0\ket{\lambda^0_1} & \leq (1-U_0)|\braket{\lambda^0_0}{\lambda^0_1}|^2 + U_0,\\ 
     \label{eq:suppdiagbounds2}\bra{\lambda^0_1}\rho_0\ket{\lambda^0_1} & \geq (1-U_0)|\braket{\lambda^0_0}{\lambda^0_1}|^2.
\end{align}
Analogous bounds can be given for $\bra{\lambda^0_0}\rho_1\ket{\lambda^0_0}$.

Finally, we need to estimate some elements of the off-diagonal blocks of $\rho_P$, or else there would be no way to differentiate an entangeld state from a classical mixture of the conditional states. To this end, we label the off-diagonal block of the full density matrix $\rho_{AB}$ by $\rho_{01}$, so that it is naturally split into the form
\begin{equation}\label{eq:rhoblocks}
 \rho_{AB} = \frac{1}{2}\begin{bmatrix}
                 \rho_{0} & \rho_{01}\\
                 \rho_{01}^{\dagger} & \rho_{1}
                \end{bmatrix},
\end{equation}
where the diagonal blocks correspond to the two conditional states. In the PM picture, we hold full knowledge of the Alice's reduced density matrix 
\begin{equation}\label{eq:rhoA}
 \rho_A = \frac{1}{2}\begin{bmatrix}
                 1 & c\\
                 c^* & 1
                \end{bmatrix},
\end{equation}
where $c=\braket{\alpha}{-\alpha}$. Each element in Eq. (\ref{eq:rhoA}) is the trace of the corresponding element in Eq. (\ref{eq:rhoblocks}), so we can enforce the condition $\tr(\rho_{01}) = c$. Using this as our starting point, and with an appropriate basis choice for system B, we can determine the following off-diagonal bounds which can be incorporated into $\rho_P$:
\begin{align}\label{eq:offdiagbound1}
 |\bra{\lambda^0_0}\rho_{01}\ket{\lambda^0_0}| \geq & |c| - \sqrt{U_0}\sqrt{1-(1-U_1)|\braket{\lambda^0_0}{\lambda^0_1}|^2},\\
 \label{eq:offdiagbound2}
 |\bra{\lambda^0_1}\rho_{01}\ket{\lambda^0_1}| \geq & |c| - \sqrt{U_1}\sqrt{1-(1-U_0)|\braket{\lambda^0_0}{\lambda^0_1}|^2}.
\end{align}
Details on how to arrive at these inequalities can be found in Appendix \ref{app:offdiagbounds}. 

We now have sufficient information to construct a useful estimate of the projected state. To summarize, we have the quantities $\kappa$ and $U_j~(j=0,1)$, which can be calculated from measurements of the first moments and second moments, respectively. We want to determine $\rho_P$, which is the projection of $\rho_{AB}$ from Eq. (\ref{eq:rhoblocks}) onto the subspace spanned by $\{\ket{0}_A,\ket{1}_A\}\otimes\{\ket{\lambda_0^0}_B,\ket{\lambda_1^0}_B\}$. We have estimated some of the overlaps of $\rho_{AB}$ with these basis vectors in Eqs. (\ref{eq:suppdiagbounds1}-\ref{eq:suppdiagbounds2}) and (\ref{eq:offdiagbound1}-\ref{eq:offdiagbound2}). These estimates depend only on the input parameter $c=\braket{\alpha}{-\alpha}$ and on the output state quantities $U_0$, $U_1$, and $|\braket{\lambda^0_0}{\lambda^0_1}|$. This last overlap quantity is itself bounded to a region defined by Eqs. (\ref{eq:overlapbounds}-\ref{eq:bupper}), which depends only on $U_0$, $U_1$ and $\kappa$. Hence, for a fixed input overlap $c$ and a fixed set of homodyne measurement results, we have a parameter region which forms a set of constraints on $\rho_P$. This region must be searched to find the minimal entanglement compatible with $\rho_P$. We will now move on to address the question of how to find the minimal entanglement compatible with our constraints. 

\subsection{Minimizing the entanglement}\label{ssec:minimize}
As mentioned earlier, we will choose the Negativity as the entanglement measure for demonstrating our method. In principle, we would like to find the minimal entanglement using the methods of semidefinite programming. But we must make some simplifications and relaxations which will allow us to do so. First, we exploit the fact that local unitary operations cannot change the quantity of entanglement. Therefore, without loss of generality, we can assume that the overlap of the maximal eigenstates is real and positive (since this can be accomplished by a relative change of phase on subsystem B).
\begin{equation}
\braket{\lambda_0^0}{\lambda_1^0} \geq 0.
\end{equation}
As well, we can perform local phase changes on subsystem A, which allows us to also make the restriction
\begin{equation}
     \bra{\lambda_0^0}\rho_{01}\ket{\lambda_0^0} \geq 0.
\end{equation}

\begin{figure}
	\includegraphics[width= 0.7 \columnwidth]{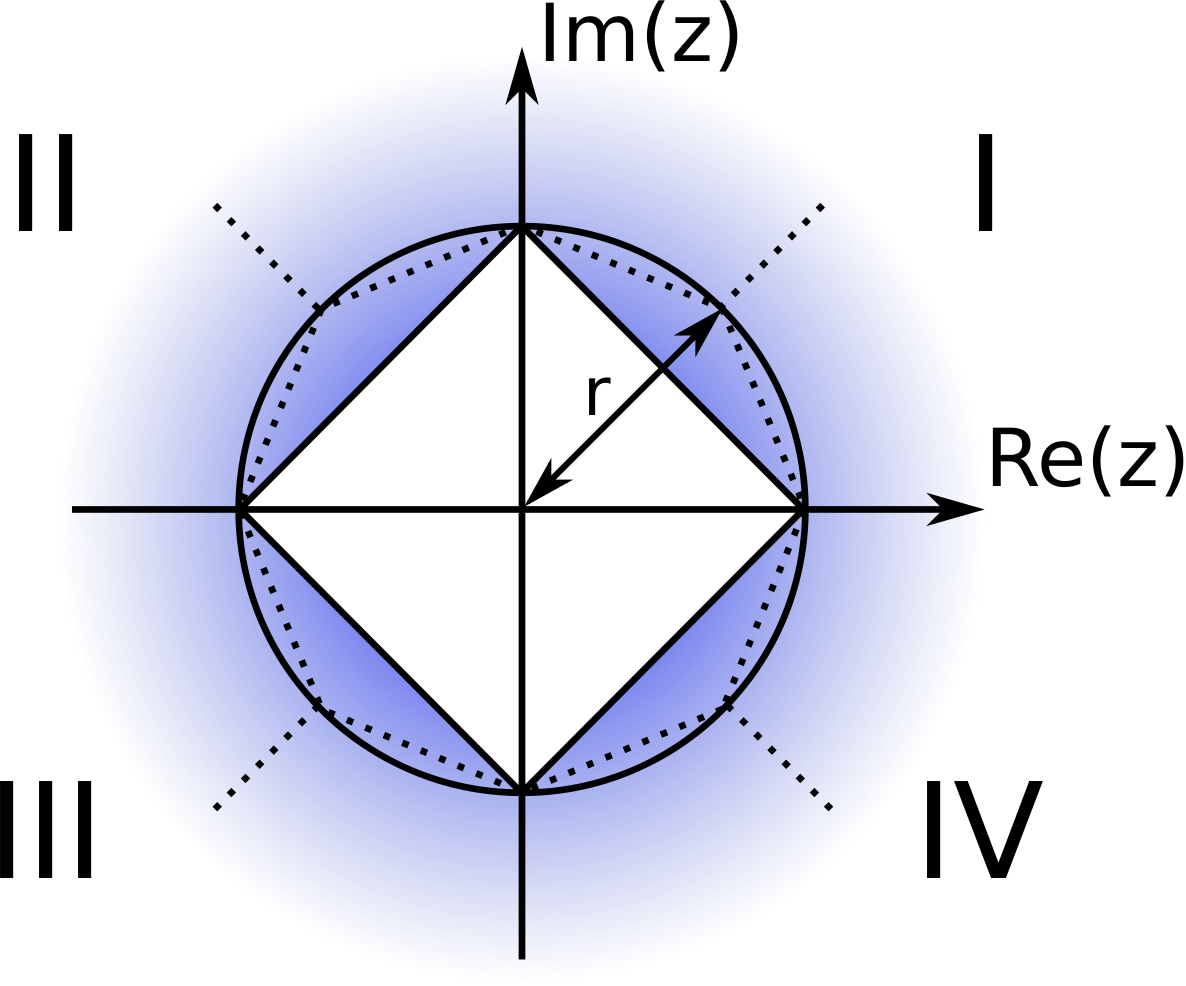}
	\caption{(Color online) Relaxing the non-convex constraint Eq. (\ref{eq:offdiagbound2}) by inscribing a square within the circle (solid lines). Four new convex search regions (I-IV) are defined by the intersection of each quadrant with the weaker constraint Eq (\ref{eq:newconstraint}). These correspond to the outer regions in the diagram. Better approximations of the circular region can be made using an inscribed octagon or other polygons, but this increases the number of convex regions that must be searched for the overall minimum (dotted lines).}
	\label{fig:convexregions}
\end{figure}

The other off-diagonal element of interest, $\bra{\lambda_1^0}\rho_{01}\ket{\lambda_1^0} =: z$, is in general still a complex number. The main problem is that Eq. (\ref{eq:offdiagbound2}) is a \emph{non-convex} constraint on $z$. To use this constraint in a semidefinite program, we have to replace it with a set of convex constraints. We accomplish this by denoting the right-hand side of Eq. (\ref{eq:offdiagbound2}) as
\begin{equation}
     |c| - \sqrt{U_1}\sqrt{1-(1-U_0)|\braket{\lambda^0_0}{\lambda^0_1}|^2} =: r
\end{equation}
and expanding our constraints to the region 
\begin{equation}\label{eq:newconstraint}
      |\text{Re}(z)|+|\text{Im}(z)| \geq r.
\end{equation}
This new constraint still non-convex, but we can search for the minimum entanglement independently in each of the four quadrants, where the constraints are convex (see Fig. \ref{fig:convexregions}), and take the minimum over these four searches. The final result will be a lower bound to the minimum entanglement in the region constrained by Eq. (\ref{eq:offdiagbound2}). We can extend this idea further, replacing the inscribed square from Fig. \ref{fig:convexregions} with any other inscribed polygon. With more sides, we can better approximate the non-convex constraint Eq. (\ref{eq:offdiagbound2}), but this will also increase the number of convex subregions which must be searched to find the overall minimum. Numerical evidence indicates that the minimum entanglement is often, though not always, found at a point outside the circle. We tested with an inscribed octagon and it was not found to alter the final results significantly. 

The final hurdle comes from the overlap $\braket{\lambda_0^0}{\lambda_1^0}$. Since the maximal eigenstates will in general have a non-zero overlap (indeed, for zero overlap, we will not find any entanglement in $\rho_P$), we must construct an orthogonal basis in order to explicitly write down a matrix representing $\rho_P$. Doing so introduces matrix elements that are both linear and quadratic in the overlap $\braket{\lambda_0^0}{\lambda_1^0}$. If the overlap is used as a parameter in the semidefinite programming, this non-linear dependence becomes problematic. Fortunately, it turns out that to find the minimal entanglement we only need to consider the case where the overlap takes the largest allowed value, i.e. $\braket{\lambda_0^0}{\lambda_1^0}=b_u$. The reason for this is that, for fixed values of $\lambda_0^0$, $\lambda_1^0$, and $\braket{\lambda_0^0}{\lambda_1^0}$, there always exists a CPTP map on the B subsystem which preserves the maximal eigenvalues while making the corresponding overlap larger. Such a local map cannot increase the entanglement, so indeed the minimal entanglement will be found at $\braket{\lambda_0^0}{\lambda_1^0}=b_u$. This useful result will be shown in detail elsewhere \cite{killoran10prep}.

\section{Results}\label{sec:results}
\begin{figure*}
  \subfloat[T=1.0]{\label{fig:test1}\includegraphics[width=0.33\textwidth]{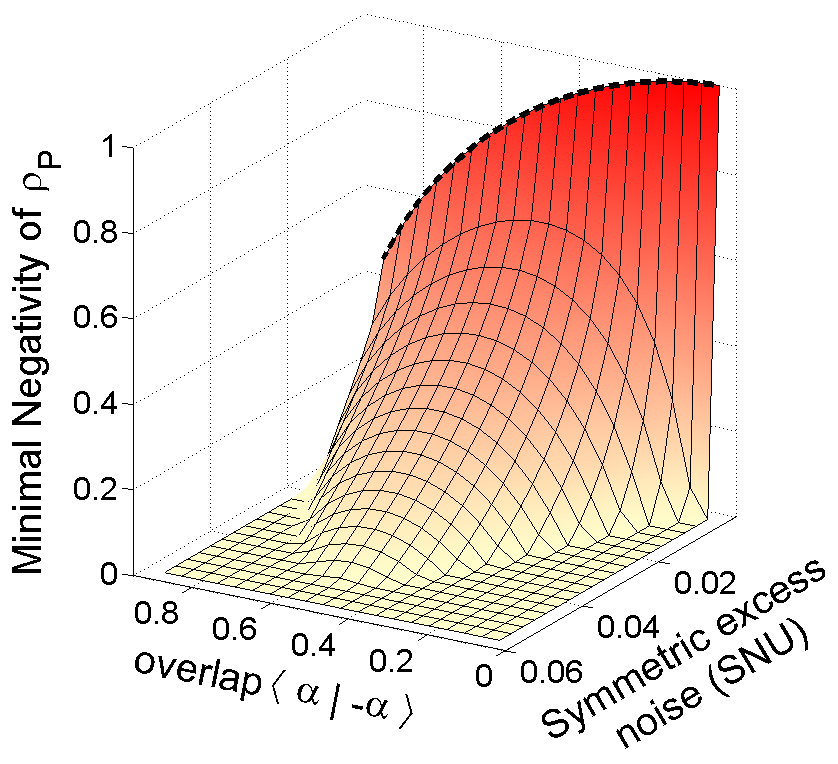}}                
  \subfloat[T=0.7]{\label{fig:test2}\includegraphics[width=0.33\textwidth]{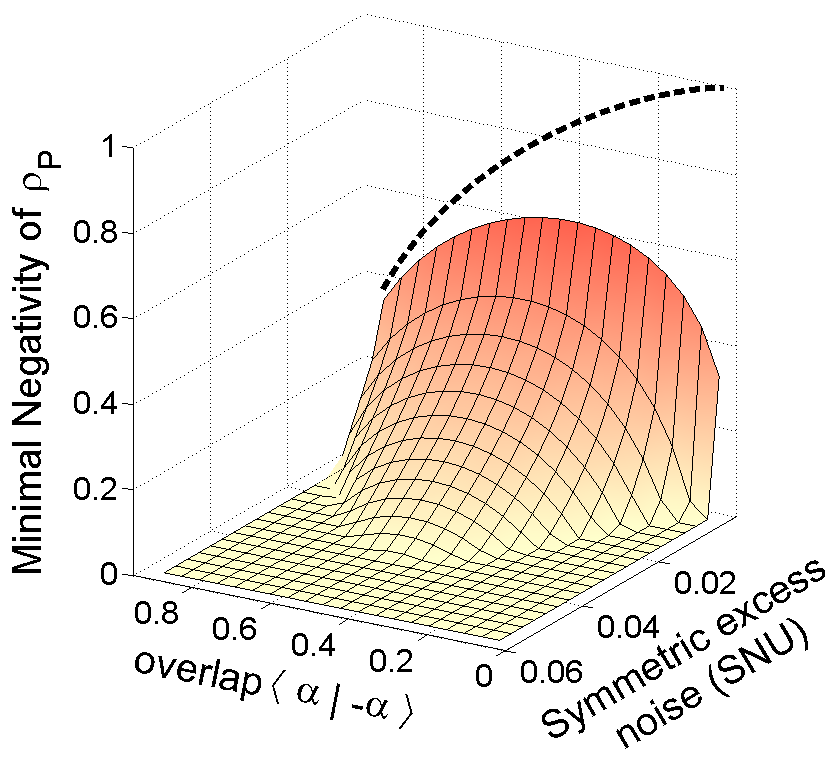}}
  \subfloat[T=0.5]{\label{fig:test3}\includegraphics[width=0.33\textwidth]{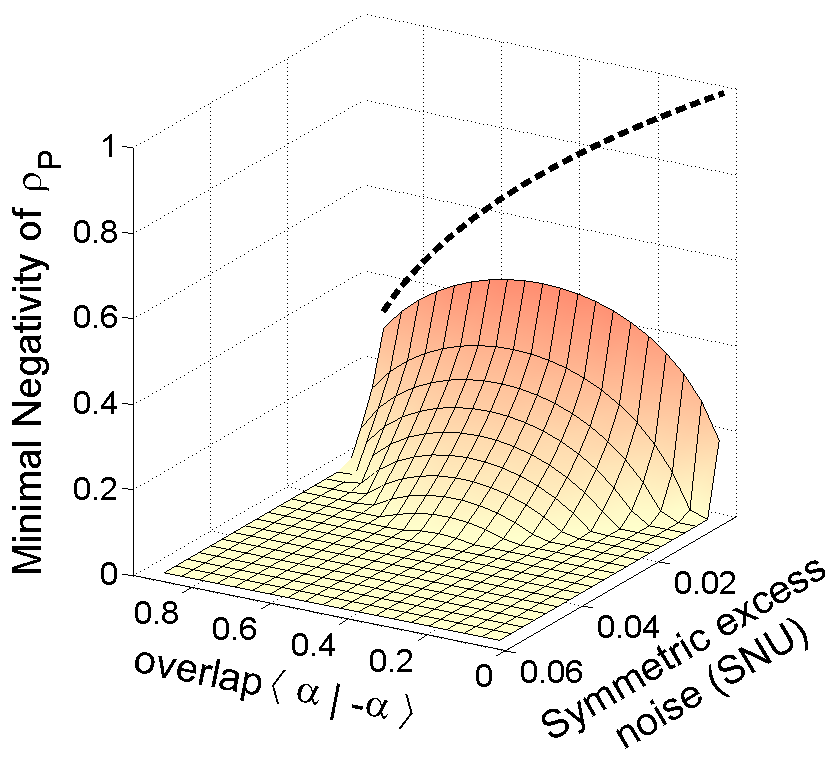}}
  \caption{(Color online) Minimal Negativity of $\rho_P$ consistent with initial overlap $\braket{\alpha}{-\alpha}$ and channel loss and noise parameters. For comparison, the dashed lines show the Negativity for states of the form from Eq. (\ref{eq:initialstate}), with $\alpha$ replaced by $\sqrt{T}\alpha$. For no loss and no noise, the bound exactly matches the initial entanglement. As either loss or noise increases, the bound lowers, until it becomes trivially equal to zero. Excess variances are expressed as a fraction of the vacuum variance (Shot Noise Units).}

	\label{fig:mainresults}
\end{figure*}

In the previous section, we outlined a method for calculating the effective entanglement in optical systems. This began with the observation that we can get bounds just by looking at the most significant two-qubit subsystem. The remainder of Sec. \ref{sec:quant} provided the necessary tools to allow us to calculate these bounds efficiently as a semidefinite program. Now that all the pieces are in place, we can turn to applying our scheme. 

To illustrate our quantification method, we use data corresponding to the action of the optical channel on the field quadratures, which we assume to be symmetric for both signal states and for both quadratures. These symmetry assumptions are made solely to aid the graphical representation of our results, and our method does not rely on them. It is also important to note that, beyond the symmetry, we do not make any assumptions about how the channel works. In the absence of experimental data, we merely parameterize the channel's effect on the first quadrature moments by a loss parameter and on the second moments by the excess noise. Specifically, if the means of the two conditional output states are denoted by $\beta_j$ from Eq. (\ref{eq:cohmean}), then the loss is parameterized through the transmittivity
\begin{equation}
     T = \frac{|\beta_0|^2}{|\alpha_0|^2} = \frac{|\beta_1|^2}{|\alpha_1|^2},
\end{equation}
and the symmetric excess noise (expressed in Shot Noise Units) by 
\begin{equation}
     V = 2\left(V_j(\hat{x})-\frac{1}{2}\right)=2\left(V_j(\hat{p})-\frac{1}{2}\right).
\end{equation}
The input states are characterized entirely by the overlap parameter $\braket{\alpha}{-\alpha}$. 

The quantification program was carried out using the Negativity \cite{zyczkowski98a,vidal02a,lee00a},
\begin{equation}
     \mathcal{N}(\rho) = ||\rho^{T_A}||_{Tr}-\tr\rho.
\end{equation}
This measure has all the properties demanded by our quantification method, but more importantly, the trace norm $||\cdot||_{Tr}$ of a matrix can be computed efficiently as a semidefinite program \cite{fazel01a}. We have normalized the Negativity so that a maximally entangled two-qubit state has $\mathcal{N}=1$. Our calculations were done in Matlab using the YALMIP interface \cite{lofberg04a} along with the solver SDPT3 \cite{toh99a}. Our main results are shown in Fig. \ref{fig:mainresults}, where the minimal Negativity of $\rho_P$ compatible with the initial overlap $\braket{\alpha}{-\alpha}$ and excess noise $V$ is given, for various values of the transmittivity $T$. This quantity gives a lower bound on the Negativity of the full state $\rho_{AB}$. The entanglement of the initial state, Eq. (\ref{eq:initialstate}), is also shown as a function of the initial overlap in Fig. (\ref{fig:test1}). For Figs. (\ref{fig:test2}-\ref{fig:test3}), the modification $\alpha\rightarrow\sqrt{T}\alpha$ is made to Eq. (\ref{eq:initialstate}) for these comparisons. The initial entanglement can be compared with the calculated bounds to help understand the quantum throughput of a device. In the limit of zero excess noise and zero loss, our entanglement bound is tight with the initial entanglement. 

Our bounds are quite high for very low noise, but they become lower as the measurement results get more noisy. At some point, a non-trivial entanglement bound can no longer be given, despite the fact that quantum correlations can still be proven for higher noise values (cf. \cite{haseler08a}). As well, for larger loss values, the tolerance for excess noise is lower, and the region where non-trivial bounds can be given becomes smaller. The exact noise value where our bounds become trivial depends on the initial overlap and on the measured loss, but the highest tolerable excess noise is around 5\% of the vacuum for $T=1.0$. This shrinks to about 3\% for a transmittivity of $T=0.5$. Though the quantification region is small, it is within the limits of current experimental technology \cite{wittmann10pc}.

Some entanglement degradation should be expected as the noise is increased, but, as mentioned earlier, entanglement can still be verified (though not previously quantified) under the same testing scenario up to much higher noise values than seen here \cite{rigas06a,haseler08a}. Thus, our bounds do not provide the full picture. The weakening of the bounds with higher noise is mainly due to the estimation procedure. Certain approximations become cruder (though still valid) as the noise increases. First, for higher noise, the conditional states become more mixed, spreading out into more of the infinite-dimensional mode Hilbert space. This leads to additional information being lost when we truncate down from $\rho_{AB}$ to $\rho_P$. Another problem stems from the bounds we use to estimate $\rho_P$. Higher noise leads to weaker bounds on the maximal eigenvalues from Eq. (\ref{eq:noisebound}), which weakens all other inequalities. 

\begin{figure}
	\includegraphics[width= 1 \columnwidth]{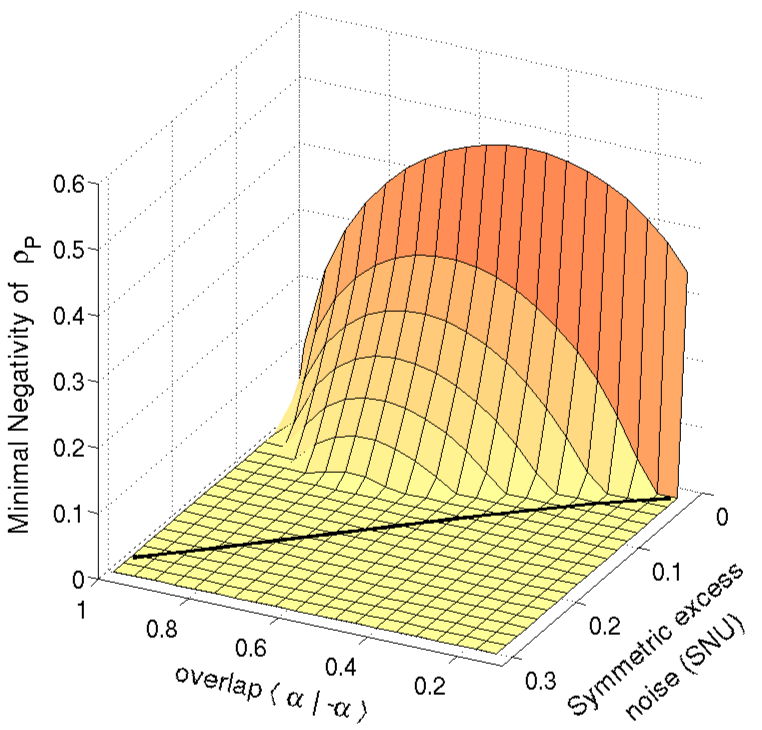}
	\caption{(Color online) Results of quantification scheme for a simple test channel where we can exactly determine the two-qubit projection parameters $\lambda_0^0$, $\lambda_1^0$, and $\braket{\lambda_0^0}{\lambda_1^0}$. The overall channel transmittivity is T=0.5, as in Fig. (\ref{fig:test3}). We also show the region where entanglement can be verified (points with lower noise than the black line). The two-qubit projection gives non-trivial entanglement bounds for roughly half the points where the entanglement can be verified.}
	\label{fig:comparison}
\end{figure}

To examine the effects of these two approximations, we briefly consider a simple channel where the test state, Eq. (\ref{eq:initialstate}), is mixed at a $50:50$ beam-splitter with a thermalized vacuum. The first moments reduce by a factor of $\frac{1}{\sqrt{2}}$, and the increased variances of the output optical states can be determined from the mean photon number $\expval{n}$ of the thermal state. For $\expval{n}>0$, the conditional output states are displaced thermal states. The reason for studying this channel is that we can \emph{exactly} determine the maximal eigenvalues $\lambda_0^0$, $\lambda_1^0$, and the overlap $\braket{\lambda_0^0}{\lambda_1^0}$. This allows us to study our approximations independently, since we decouple the effects of the two-qubit projection from the homodyne parameter estimation (in practice, of course, our quantification scheme must use both). In Fig. (\ref{fig:comparison}) we show the result of the quantification scheme, when this extra information is included. We see that the tolerable excess noise is $>10\%$ of the vacuum, more than three times what it would be if we had to estimate the eigenvalues and overlap using homodyne results (cf. Fig. (\ref{fig:test3})). Also included in Fig. (\ref{fig:comparison}) is an entanglement verification curve, obtained using the methods of \cite{rigas06a,haseler08a}. Any points with lower noise than this verification curve must come from entangled states. The two-qubit projection is tight to the entanglement verification curve for low overlaps. For higher values, the projection becomes weaker, only working to about half the noise value that the entanglement verification curve reaches. 

Ideally, we want to be able to calculate non-trivial values for the entanglement wherever it be verified. This would give us a true quantitative complement to existing entanglement verification methods. One obvious extension to our method would be to truncate the mode subspace using the two largest eigenstates from each conditional state, or even more. In theory, this would strictly improve the estimates. However, in practice, this will increase the complexity of the quantification calculation, since some simplifying assumptions (i.e. certain overlaps are real) may no longer be valid. As well, the number of additional minimizations we have to do, as in our non-convex relaxation of Eq. (\ref{eq:offdiagbound2}), increases fourfold with each added dimension. Another approach might therefore be necessary to overcome this problem. Nevertheless, the quantification scheme outlined here is a useful method for characterizing the degree of quantumness of optical channels, especially when these channels introduce low noise.

\section{Conclusion}\label{sec:conclusion}
We have outlined a method for quantifying the effective entanglement in qubit-mode systems using only homodyne measurement results and knowledge of the initial preparation. This quantification method works particularly well if the mode subsystem exhibits low noise. By combining this quantification scheme with a device testing scenario which uses two nonorthogonal test states, one can examine how strongly an optical device or experiment is operating in the quantum domain. Our scheme provides a useful tool for understanding the quantum nature of optical devices, especially the question of how well they preserve quantum correlations.

\acknowledgments{Portions of this work were carried out at the National Institute for Informatics in Tokyo, Japan. As well, this work was finished while N.K. was visiting the Max Planck Institute for the Science of Light in Erlangen, Germany, and he is grateful for support from the Collaborative student training in Quantum Information Processing program. Support from NSERC (Discovery Grant and Quantum Works) and OCE are also acknowledged.}

\appendix

\section{Derivation of overlap bounds}\label{app:overlapbounds}
In this appendix, we derive the bounds from Eqs. (\ref{eq:blower}-\ref{eq:bupper}) for the absolute value of the overlap of the maximal eigenstates, $|\braket{\lambda_0^0}{\lambda_1^0}|$. From \cite{rigas06b,zhao09a}, we have the following:

\emph{Overlap Bounds.} Let the largest eigenvalue of $\rho_j$ be parameterized by 
\begin{equation}
 \lambda_j^0 =: 1-\tilde{\varepsilon}_j,
\end{equation}
and let the fidelity between the conditional states and the coherent states $\ket{\beta_j}$ from Eq. (\ref{eq:cohmean}) be given by 
\begin{equation}
 \bra{\beta_j}\rho_j\ket{\beta_j} =: 1-\varepsilon_j
\end{equation}
and let 
\begin{equation}
 \kappa := |\braket{\beta_0}{\beta_1}|. 
\end{equation}
Then the following holds:
\begin{equation}
 c_l(\kappa, \varepsilon_j, \tilde{\varepsilon}_j) \leq |\braket{\lambda_0^0}{\lambda_1^0}| \leq c_u(\kappa, \varepsilon_j, \tilde{\varepsilon}_j),
\end{equation}
with
\begin{align}
      	c_l = & ~\displaystyle\kappa\sqrt{\frac{1-\varepsilon_0-\tilde{\varepsilon}_0}{1-2\tilde{\varepsilon}_0}}\sqrt{\frac{1-\varepsilon_1-\tilde{\varepsilon}_1}{1-2\tilde{\varepsilon}_1}}\notag\\
	      & -\displaystyle\sqrt{1-\kappa^2}\sqrt{\frac{1-\varepsilon_0}{1-\tilde{\varepsilon}_0}}\sqrt{\frac{\varepsilon_1-\tilde{\varepsilon}_1}{1-2\tilde{\varepsilon}_1}}\notag\\
	      & -\displaystyle\sqrt{1-\kappa^2}\sqrt{\frac{1-\varepsilon_1}{1-\tilde{\varepsilon}_1}}\sqrt{\frac{\varepsilon_0-\tilde{\varepsilon}_0}{1-2\tilde{\varepsilon}_0}}\notag\\	
	      & -\displaystyle\sqrt{\frac{\varepsilon_1-\tilde{\varepsilon}_1}{1-2\tilde{\varepsilon}_1}}\sqrt{\frac{\varepsilon_0-\tilde{\varepsilon}_0}{1-2\tilde{\varepsilon}_0}}
	      \label{eq:oldoverlaplowerbound}
\end{align}
and
\begin{align}
      	c_u = & ~\displaystyle\kappa\sqrt{\frac{1-\varepsilon_0}{1-\tilde{\varepsilon}_0}}\sqrt{\frac{1-\varepsilon_1}{1-\tilde{\varepsilon}_1}}\notag\\
	      & +\displaystyle\sqrt{1-\kappa^2}\sqrt{\frac{1-\varepsilon_0}{1-\tilde{\varepsilon}_0}}\sqrt{\frac{\varepsilon_1-\tilde{\varepsilon}_1}{1-2\tilde{\varepsilon}_1}}\notag\\
	      & +\displaystyle\sqrt{1-\kappa^2}\sqrt{\frac{1-\varepsilon_1}{1-\tilde{\varepsilon}_1}}\sqrt{\frac{\varepsilon_0-\tilde{\varepsilon}_0}{1-2\tilde{\varepsilon}_0}}\notag\\	
	      & +\displaystyle\sqrt{\frac{\varepsilon_1-\tilde{\varepsilon}_1}{1-2\tilde{\varepsilon}_1}}\sqrt{\frac{\varepsilon_0-\tilde{\varepsilon}_0}{1-2\tilde{\varepsilon}_0}}.
	      \label{eq:oldoverlapupperbound}
\end{align}

Since we cannot calculate $\varepsilon$ or $\tilde{\varepsilon}$ in practice, we now modify these bounds from the above form found in \cite{rigas06b,zhao09a} to one involving only the parameters $\kappa$ (calculated from first moments) and the $U_j$ (calculated from second moments). To do this, we make use only of the obvious inequality 
\begin{equation}
 0\leq\tilde{\varepsilon}_j\leq \varepsilon_j\leq U_j.
\end{equation}
From this, we can easily derive the following auxiliary inequalities:
\begin{align}
     &\displaystyle\sqrt{\frac{1-\varepsilon_j}{1-\tilde{\varepsilon_j}}}\leq 1\label{eq:aux1}\\
     &\displaystyle\sqrt{\frac{\varepsilon_j-\tilde{\varepsilon_j}}{1-2\tilde{\varepsilon}_j}} \leq \sqrt{\frac{\varepsilon_j}{1-2\tilde{\varepsilon}_j}} \leq \sqrt{\frac{U_j}{1-2U_j}}, ~ \left(U_j<\frac{1}{2}\right)\label{eq:aux2}\\
     &\displaystyle\sqrt{\frac{1-\varepsilon_j-\tilde{\varepsilon}_j}{1-2\tilde{\varepsilon}_j}} \geq \sqrt{1-2U_j}, ~ \left(U_j<\frac{1}{2}\right).\label{eq:aux3}
\end{align}
It is important to note that the second and third inequalities only hold so long as $U<\frac{1}{2}$. For symmetric noise, the value $U = \frac{1}{2}$ corresponds to $V_j(\hat{x}) = V_j(\hat{p}) = \sqrt{2}-\frac{1}{2}\approx0.914$, almost twice the vacuum variance. This value is far outside the region where our method gives non-trivial bounds, so it is not an issue. Substituting the inequalities (\ref{eq:aux1}-\ref{eq:aux3}) into Eqs. (\ref{eq:oldoverlaplowerbound}) and (\ref{eq:oldoverlapupperbound}), we arrive at the bounds given in Eqs. (\ref{eq:blower}-\ref{eq:bupper}).

\section{Derivation of bounds on supplementary fidelities}\label{app:suppdiagbounds}
Here we aim to bound the quantities $\bra{\lambda_i^0}\rho_j\ket{\lambda_i^0}$ for $i\neq j$, as found in Eqs. (\ref{eq:suppdiagbounds1}-\ref{eq:suppdiagbounds2}). An eigenbasis expansion of $\rho_0$ leads to 
\begin{align}
	\bra{\lambda_1^0}\rho_0\ket{\lambda_1^0} = & ~(1-\tilde{\varepsilon}_0)|\braket{\lambda_0^0}{\lambda_1^0}|^2 + \sum_{k=1}^{\infty}\lambda_0^k|\braket{\lambda_1^0}{\lambda_0^k}|^2\notag\\
	 \leq & ~(1-\tilde{\varepsilon}_0)|\braket{\lambda_0^0}{\lambda_1^0}|^2 + \sum_{k=1}^{\infty}\lambda_0^k\notag\\
	 = & ~(1-\tilde{\varepsilon}_0)|\braket{\lambda_0^0}{\lambda_1^0}|^2 + \tilde{\varepsilon}_0\notag\\
	 = & ~\tilde{\varepsilon}_0 (1-|\braket{\lambda_0^0}{\lambda_1^0}|^2) + |\braket{\lambda_0^0}{\lambda_1^0}|^2\notag\\ 
	 \leq & ~U_0 (1-|\braket{\lambda_0^0}{\lambda_1^0}|^2) + |\braket{\lambda_0^0}{\lambda_1^0}|^2\notag\\
	 = & ~(1-U_0)|\braket{\lambda_0^0}{\lambda_1^0}|^2 + U_0.
\end{align}
A lower bound can be derived in a similar way:
\begin{align}
	\bra{\lambda_1^0}\rho_0\ket{\lambda_1^0} = & (1-\tilde{\varepsilon}_0)|\braket{\lambda_0^0}{\lambda_1^0}|^2 + \sum_{k=1}^{\infty}\lambda_0^k|\braket{\lambda_1^0}{\lambda_0^k}|^2\notag\\
	\geq & (1-\tilde{\varepsilon}_0)|\braket{\lambda_0^0}{\lambda_1^0}|^2\notag\\
	\geq & (1-U_0)|\braket{\lambda_0^0}{\lambda_1^0}|^2.
\end{align}
The bounds for $\bra{\lambda_0^0}\rho_1\ket{\lambda_0^0}$ follow by interchanging indices.

\section{Derivation of off-diagonal bounds}\label{app:offdiagbounds}
This appendix outlines the derivation of the off-diagonal bounds from Eqs. (\ref{eq:offdiagbound1}-\ref{eq:offdiagbound2}). We completely know $\rho_A$, which constrains that we must have $\text{Tr}\rho_{01} = \braket{\alpha}{-\alpha} = c$. First, we consider the full density matrix $\rho_{AB}$ in the basis defined by $\{\ket{0},\ket{1}\}$ for system $A$ and the eigenbasis of $\rho_0$, $\{ \ket{\lambda_0^k} \}_{k=0}^{\infty}$, for system B. We can still write this in the block form of Eq. (\ref{eq:rhoblocks}), where we denote the diagonal elements of the block $\rho_1$ by $\{b_k\}_{k=0}^{\infty}$ and the diagonal elements of the block $\rho_{01}$ by $\{d_k\}_{k=0}^{\infty}$ (the diagonal elements of $\rho_0$ are its eigenvalues). Using the triangle inequality, we have
\begin{equation}
     |c| = |\text{Tr}\rho_{01}| = \left| d_0 + \sum_{k=1}^{\infty}d_k\right| \leq |d_0| + \sum_{k=1}^{\infty}|d_k|.
\end{equation}
From positivity of $\rho_{AB}$, we find 
\begin{equation}
     |c|\leq |d_0| + \sum_{k=1}^{\infty}\displaystyle\sqrt{\lambda_0^k}\displaystyle\sqrt{b_k},
\end{equation}
and from the Cauchy-Schwarz inequality,
\begin{equation}
     |c| \leq |d_0| + \sqrt{\left(\sum_{k=1}^{\infty}\lambda_0^k\right)\left(\sum_{k=1}^{\infty}b_k\right)}.
\end{equation}
The first sum is just $\tilde{\varepsilon}_0$ and the second is $1-b_0$. Now, using the bounds from Appendix \ref{app:suppdiagbounds}, we get 
\begin{align}
	b_0 = & \bra{\lambda_0^0}\rho_1\ket{\lambda_0^0}\notag\\
	 \geq & (1-U_1)|\braket{\lambda_0^0}{\lambda_1^0}|^2,
\end{align}
which we can substitute above to obtain
\begin{equation}
|c| \leq |d_0| + \sqrt{\tilde{\varepsilon}_0}\sqrt{1-(1-U_1)|\braket{\lambda_0^0}{\lambda_1^0}|^2}.
\end{equation}
Replacing $d_0$ with $\bra{\lambda_0^0}\rho_{01}\ket{\lambda_0^0}$, we are led to the off-diagonal bound
\begin{equation}
     |\bra{\lambda_0^0}\rho_{01}\ket{\lambda_0^0}| \geq |c| - \sqrt{U_0}\sqrt{1-(1-U_1)|\braket{\lambda_0^0}{\lambda_1^0}|^2}.
\end{equation}
By applying the same arguments using the eigenbasis of $\rho_1$, we can arrive at an analogous bound for $|\bra{\lambda_1^0}\rho_{01}\ket{\lambda_1^0}|$.

\end{document}